\documentclass[prl,showpacs,twocolumn]{revtex4}
\usepackage{latexsym}
\usepackage{amssymb}
\usepackage{amsmath}
\usepackage{graphicx}
\usepackage{subfigure}

\begin{document}

\title{Probing Quantum Hall Pseudospin Ferromagnet by Resistively Detected
NMR }
\author{G. P. Guo$^{(1)}$}
\author{Y. J. Zhao$^{(1)}$}
\author{T. Tu$^{(1)}$}
\email{tutao@ustc.edu.cn}
\author{X. J. Hao$^{(1)}$}
\author{G. C. Guo$^{(1)}$}
\author{H. W. Jiang$^{(2)}$}
\email{jiangh@physics.ucla.edu}
\affiliation{$^{(1)}$ Key Laboratory of Quantum Information, University of Science and
Technology of China, Chinese Academy of Sciences, Hefei 230026, P. R. China\\
$^{(2)}$ Department of Physics and Astronomy, University of California at
Los Angeles, 405 Hilgard Avenue, Los Angeles, CA 90095, USA }
\date{\today }

\begin{abstract}
Resistively Detected Nuclear Magnetic Resonance (RD-NMR) has been used to
investigate a two-subband electron system in a regime where quantum Hall
pseudo-spin ferromagnetic (QHPF) states are prominently developed. It
reveals that the easy-axis QHPF state around the total filling factor $\nu
=4 $ can be detected by the RD-NMR measurement. Approaching one of the
Landau level (LL) crossing points, the RD-NMR signal strength and the
nuclear spin relaxation rate $1/T_{1}$ enhance significantly, a signature of
low energy spin excitations. However, the RD-NMR signal at another identical
LL crossing point is surprisingly missing which presents a puzzle.
\end{abstract}

\maketitle
\date{\today }


The multi-component electron systems have been continuously drawing
intensive research interest because of its novel ground states and
excitations \cite{DasSarma}. In experimental systems, different Landau
levels (LLs) can be tuned to cross by varying gate voltage, charge density,
magnetic field or the magnetic field tilted angle to the sample.
Electron-electron correlations become particularly prominent when two or
more sets of LLs with different layer, subband, valley, spin, or Landau
level indices are brought into degeneracy \cite%
{DasSarma,Wescheider1999,Shayegan2000,Hirayama2001,Jiang2005,Jiang2006,Tsui2006,Shayegan2006}%
. Recent experiments in single quantum well with two subbands occupied
systems \cite{Jiang2005,Jiang2006}, showed evidence of the formation of
quantum Hall pseudospin ferromagnets (QHPFs) due to the interactions of the
two subbands (termed as pseudospins) around the LLs crossing point. The
QHPFs taking place at total filling factor $\nu =3,5$ and $\nu =4$ are
easy-plane or easy-axis QHPFs respectively, depending on the details of the
two subbands configurations. In spite of various theoretical models \cite%
{MacDonald2000,DasSarma2003,Hao2008} motivated by these findings, a
comprehensive understanding is not yet achieved. Thus far, experimental and
theoretical studies all focused on the pseudospin freedom. However, in this
work we would address the unique spin excitations in the QHPF states.

To address the question whether spin states in two-subband systems in
nature, measurements other than the conventional transport and optical means
are needed. Since the Zeeman energy of nuclear spin is about $3$ orders of
magnitude smaller than that of electron spin, exchange of spin angular
momentum between the electron and nuclear spin is allowed only when the
electron system supports spin excitations with low energy. The nuclear spin
relaxation rate $1/T_{1}$ thus probes the density of states at low energy of
the electron spin system that cannot be accessed by other means. The
resistively detected NMR technique has recently emerged as an effective
method to probe collective spin states in the fractional quantum Hall regime
\cite{KlitzingNMR1,KlitzingNMR2}, the Skyrmion spin texture close to the
filling factor $1$ \cite{PortalNMR,TsuiNMR}, the role of electron spin
polarization in the phase transition of a bilayer system \cite%
{EisensteinNMR,HirayamaNMR1}, and the ferromagnetic state accompanied by
collective spin excitations of a two-subband system \cite{JiangNMR}. Here we
use this technique to study spin freedom and its relation with pseudospin in
the vicinity of the QHPF states at filling factor $\nu =3,4,5$. It reveals
that the easy-axis QHPF state at $\nu =4$ is sensitive to the RD-NMR
measurement. As approaching to one LL crossing point at $\nu =4$ where the
easy-axis QHPF phase is well developed, the RD-NMR signal strength and the
nuclear spin relaxation rate $1/T_{1}$ enhance quickly which may be due to
the low energy spin excitations there. Furthermore, the RD-NMR signal can be
suppressed anomaly at another identical LL crossing point of $\nu =4$.

The sample was grown by molecular-beam epitaxy and consists of a symmetrical
modulation-doped $24$ nm wide single GaAs quantum well bounded on each side
by Si $\delta $-doped layers of AlGaAs with doping level $n_{d}=10^{12}$ cm$%
^{-2}$. Heavy doping creates a very dense 2DEG, resulting in the filling of
two subbands in the well. As determined from the Hall resistance data and
Shubnikov-de Haas oscillations in the longitudinal resistance, the total
density is $n=8.0\times 10^{11}$ cm$^{-2}$, where the first and the second
subband have a density of $n_{1}=6.1\times 10^{11}$ cm$^{-2}$ and $%
n_{2}=1.9\times 10^{11}$ cm$^{-2}$. The sample has a low-temperature
mobility $\mu =4.1\times 10^{5}$ cm$^{2}$/V s, which is extremely high for a
2DEG with two filled subbands. A $100$ $\mu $m wide Hall bar with $270$ $\mu
$m between voltage probes was patterned by standard lithography techniques.
A NiCr top gate was evaporated on the top of the sample, approximately $350$
nm away from the center of the quantum well. By applying a negative gate
voltage on the NiCr top gate, the electron density can be varied
continuously. Several turns of NMR coil were wound around the sample, which
was placed in a Top-Loading Dilution Refrigerator with a base temperature of
$15$ mK. A small radio frequency (rf) magnetic field generated by the coil
with a matching frequency $f=\gamma H_{0}$ will cause NMR for $^{75}$As
nuclei, where the gyromagnetic ratio $\gamma =7.29$ MHz/T. The resistance
was measured using quasi-dc lock-in technique with $11.3$ Hz.

\begin{figure}[tbp]
\subfigure[] {\includegraphics[width=0.8\columnwidth]{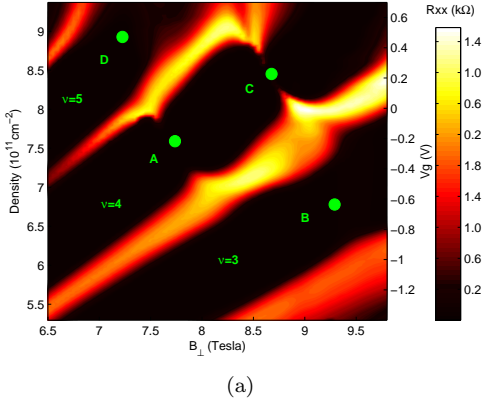}} %
\subfigure[] {\includegraphics[width=0.8%
\columnwidth]{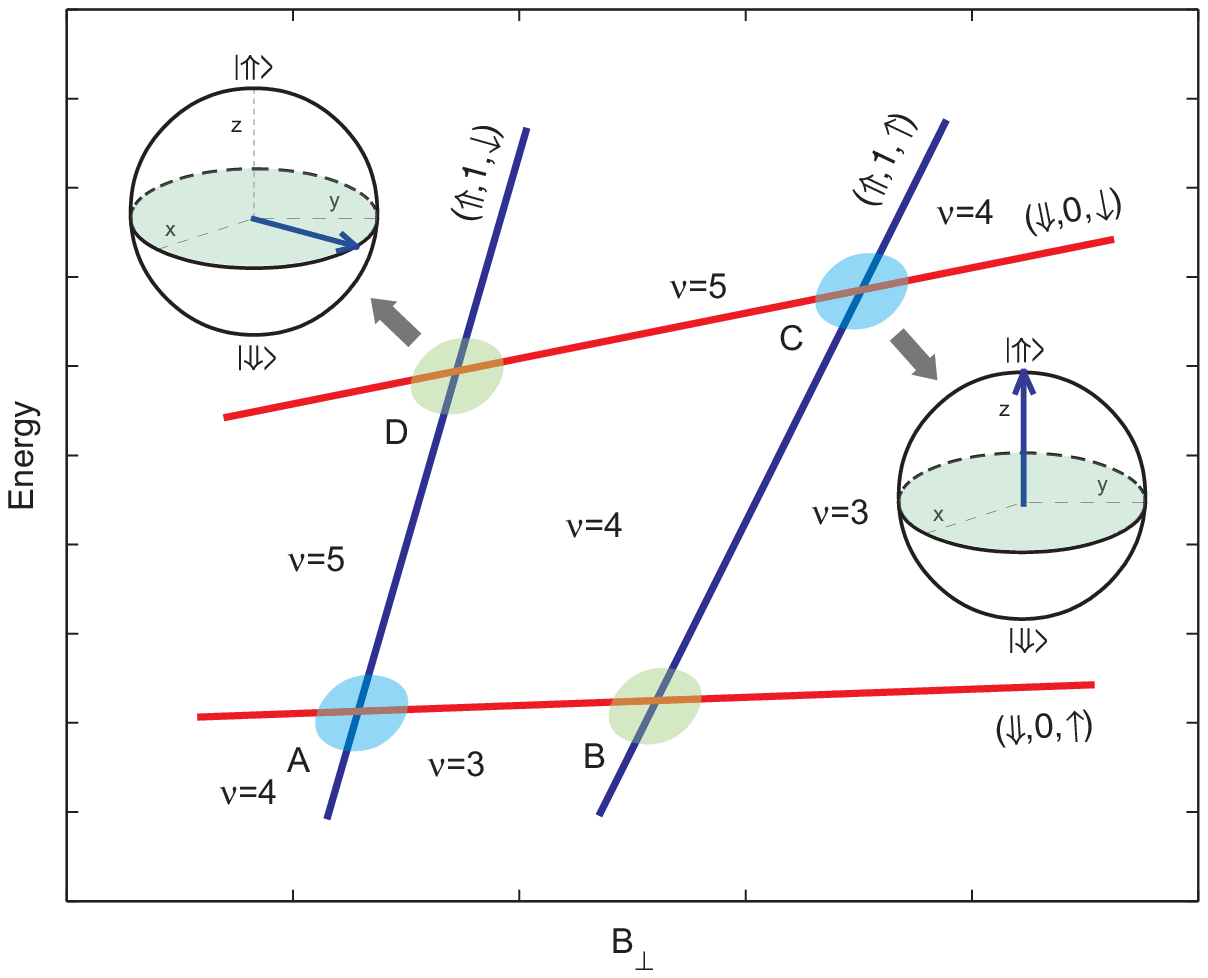}}
\caption{(a) The longitudinal resistance $R_{xx}$ in the density ($n$) -
magnetic field ($B_{\bot }$) phase diagram at filling factor $\protect\nu %
=3,4,5$, which are measured at the base temperature. (b) Schematic drawing
of the crossing between different indices Landau levels and resulting
easy-plane or easy-axis pseudo-spin states at points B, D and A, C, as
correspondingly marked in Fig. 2a. }
\label{R_Phase}
\end{figure}
\begin{figure}[tbp]
\subfigure[]{\includegraphics[width=0.8\columnwidth]{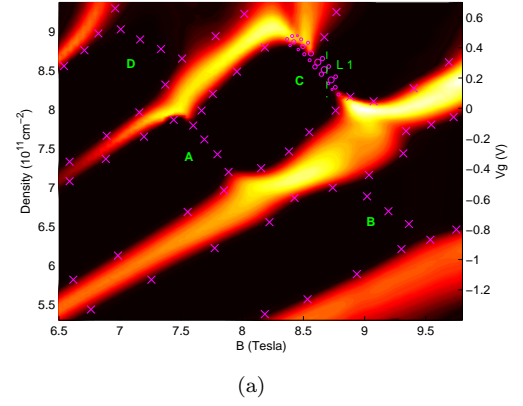}} %
\subfigure[]{\includegraphics[width=0.8\columnwidth]{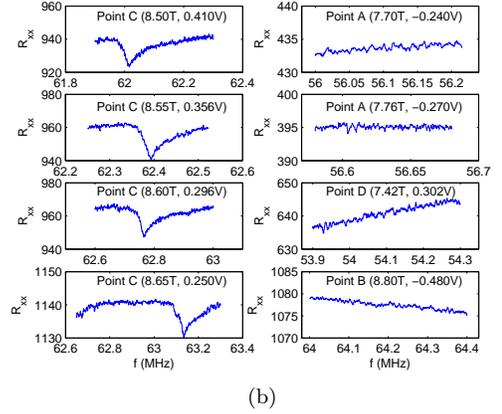}}
\caption{(a) The NMR signals phase diagram of the sample at $\protect\nu %
=3,4,5$. The cross and circle symbols in the map denote the places where the
NMR signals are measured. The '$\times $' mean places where there are no NMR
signals, while the '$\circ $' show the places where the NMR signals are
observed. And the size of '$\circ $' symbols give a schematic illustration
of the strength of NMR signals. The dashed line L1 is the trace along which
we measured NMR signal as shown in Fig. 5. (b) Typical resistively detected
NMR spectrum measured around point C and A, B, D. }
\label{NMR_Phase}
\end{figure}

In the present work, we refer the first and second subbands, to as symmetric
and antisymmetric states. In the pseudo-spin language, one of them can be
labeled as pseudo-spin up ($\Uparrow $) and the other as pseudo-spin down ($%
\Downarrow $). When a magnetic field $B_{\bot }$ is applied, the energy
spectrum of the quantum well discretizes into a sequence of Landau levels.
We label the single-particle levels ($i,N,\sigma $), which $i$ ($=\Uparrow
,\Downarrow $), $N$, and $\sigma $ ($=\uparrow ,\downarrow $) are the
pseudo-spin, orbital and spin quantum numbers. In the present work we have
concentrated our study around the filling factor $\nu =3,4,5$, where the
filling factor $\nu $ denotes the number of filled Landau levels. The
longitudinal resistance $R_{xx}$ in the density ($n$) - perpendicular
magnetic field ($B_{\bot }$) plane exhibits a square-like structure around $%
\nu =3,4,5$, as shown in Fig. \ref{R_Phase}a. The most noticeable feature of
the square-like structure is the disappearance of the extended states (i.e.,
bright lines) on its four boundaries, marked by A, B, C, D in Fig. \ref%
{R_Phase}a. Here point A corresponds to the degeneracy point of $\left\vert
(\Uparrow ,1,\downarrow )\right\rangle $ and $\left\vert (\Downarrow
,0,\uparrow )\right\rangle $, point B corresponds to that of $\left\vert
(\Uparrow ,1,\uparrow )\right\rangle $ and $\left\vert (\Downarrow
,0,\uparrow )\right\rangle $, point C corresponds to that of $\left\vert
(\Uparrow ,1,\uparrow )\right\rangle $ and $\left\vert (\Downarrow
,0,\downarrow )\right\rangle $, point D corresponds to that of $\left\vert
(\Uparrow ,1,\downarrow )\right\rangle $ and $\left\vert (\Downarrow
,0,\downarrow )\right\rangle $, as illustrated schematically in the Landau
level fan diagram Fig. \ref{R_Phase}b. The disappearance and result square
structure represents a pseudo-spin ferromagnet, which is due to the opening
pseudo-spin gaps of easy-plane or easy-axis pseudo-spin ferromagnetic
states, respectively at the level crossing points of B, D and A, C, as
depicted in Fig. \ref{R_Phase}b \cite%
{Jiang2005,Jiang2006,Hirayama2001,MacDonald2000}.

RD-NMR, performed in the proximity of the square structure, reveals
prominent (absent) NMR signal at different regions. In order to get a clear
signal and minimize heat effect, most of experiments were carried out with a
rf power of $0$ dBm. The ac current $I_{ac}$ was $50$ nA, and a large dc
current $I_{dc}=250$ nA were applied to enhance the NMR signal. All the
measurements were carried out at temperature below $120$ mK. The measurement
result under the same condition are shown in Fig. \ref{NMR_Phase}a, the
cross and circle symbols in the map denote the places where the NMR signals
are measured. The cross '$\times $' means the places where there are no NMR
signals, while the circle '$\circ $' shows the places where the NMR signals
are observed. And the size of '$\circ $' symbols give a schematic
illustration of the strength of NMR signals. From this map we found that the
NMR signals only occur at the upper arm of the square structure around
crossing point C, while we didn't find any signal at the lower arm of this
square structure around another crossing point A and its two sides around
crossing point B and D.

Now we focused on the region around the LL crossing point C, where
pronounced NMR signals were observed. Typical NMR lines around point C are
shown in Fig. \ref{NMR_Phase}b. The relative change of $R_{xx}$ is typically
about 1\% at resonance. Upon resonance, $R_{xx}$ in all NMR lines shows a
sharp decrease followed by a much slower relaxation process back to its
original value, which is characterized by the nuclear spin relaxation time
owing to the interaction with the electron spin system, $T_{1}$, as will be
discussed below. In these experiments, we have changed the rf amplitude from
$-15$ dBm to $2$ dBm. Even very weak, the NMR signal can be recognized at $%
-15$ dBm.

We believe the RD-NMR described here is due to the electron and nuclear spin
flip-flop effect \cite{JiangNMR}. For the two dimensional electron system in
GaAs, the contact hyperfine interaction with the polarized nuclei acts as an
effective magnetic field $B_{N}$ for the electron spin. The effective
electron spin-flip energy is then reduced, $E_{z}=g^{\ast }\mu
_{B}BS_{z}+A\left\langle I_{z}\right\rangle S_{z}=g^{\ast }\mu
_{B}(B+B_{N})S_{z}$ as $g^{\ast }<0$. When the NMR resonance condition is
matched, the nuclear spins are depolarized and the electron Zeeman energy
increases consequently. Since $R_{xx}$ is dependent on the thermally
activated energy gap $E_{a}$, $R_{xx}\propto \exp (-E_{a}/2k_{B}T)$, the NMR
is manifested by a drop in $R_{xx}$, as shown by all the NMR lines in Fig. %
\ref{NMR_Phase}b. This allows the nuclear spin polarization to be
sensitively detected by a change in the transport coefficient of the
electron system $R_{xx}$.

The above observations reveals the spin excitation in the square structure
is of intrinsic interest and is well correlated with the spin excitations of
the easy-axis QHPF states. At point C, when the two competing pseudospin (up
and down) states acquire the same energy and leads to easy-axis anisotropy,
they separate into domains with opposite pseudospin states \cite%
{Hirayama2001,Jiang2006,MacDonald2000,MacDonald2001}. On the other hand, the
pseudospin up and down states have opposite spins. As a result, magnetic
domains form and the electronic state within each domain is described as an
Ising-like QH ferromagnet with either one of two possible spin orientations.
As the applied current forces electrons to scatter between adjacent domains
with different spin but almost degenerate energy, the nuclei in the
neighborhood can become polarized and probed by the RD-NMR measurement.
However at other crossing point B and D, the QHPF states are easy-plane,
which means that the two degenerate Landau levels are mixing and no spin
magnetization formation. Since easy-plane QHPF state can not spontaneously
separate into magnetic domains, there is no nuclear polarization and the NMR
signals are destroyed.

To support the mechanism of the polarized nuclear spins, current dependence
of the NMR signal was studied. In this measurement, the sample resistance
was measured with a low ac current of $20$ nA, while ramping the dc current
in a wide range to bias the sample. The result indicates that the NMR signal
is enhanced by a factor of $8$ in the low current range from $100$ nA up to $%
250$ nA. The data thus consist with the picture of current induced dynamic
polarization.

\begin{figure}[tbp]
\includegraphics[width=0.8\columnwidth]{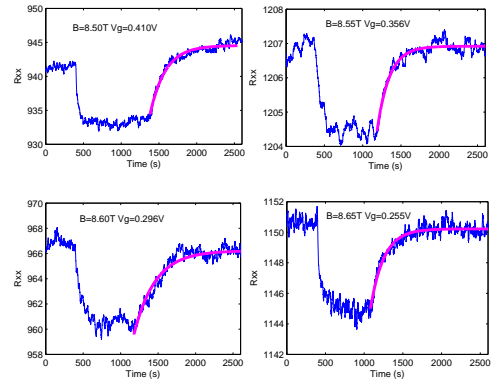}
\caption{Measuring nuclear spin relaxation time $T_{1}$ around point C by
recording time evolution of $R_{xx}$ irradiated by rf, initially off
resonance, on resonance and finally off resonance. $T_{1}$ is determined by
an exponential fit to the experiment data.}
\label{C_NMR_T1}
\end{figure}
\begin{figure}[tbp]
\subfigure[] {\includegraphics[width=0.8\columnwidth]{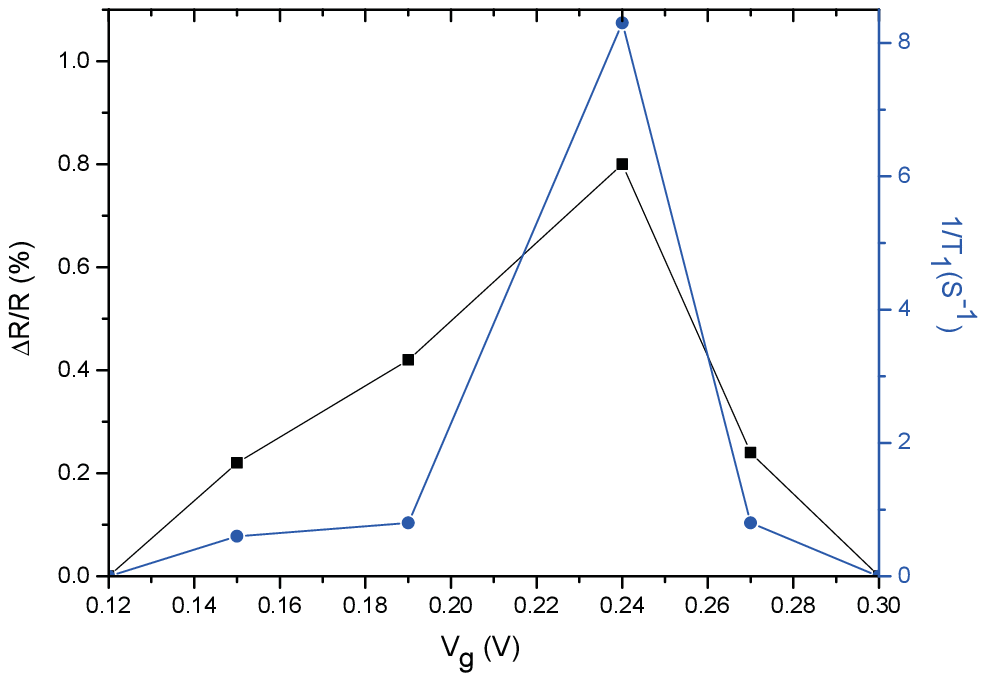}} %
\subfigure[] {\includegraphics[width=0.8\columnwidth]{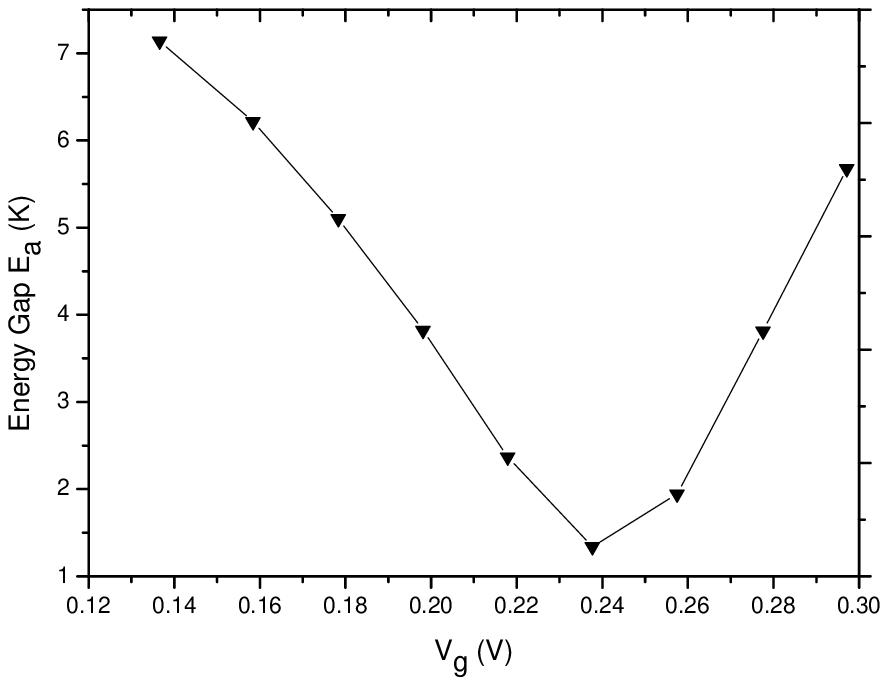}}
\caption{(a) Plot of the resistively detected NMR signal ratio $\Delta
R_{xx}/R_{xx}$ (black square), nuclear spin relaxation rate $1/T_{1}$ (blue
circle) against gate voltage $V_{g}$ along the line L1 (in Fig. 2a). (b)
Plot of electron activation energy gap $E_{a}$ against gate voltage $V_{g}$
along the same line. }
\label{L1_NMR}
\end{figure}

To gain more support of our observation of the nature of the spin in the
easy-axis QHPF states, we studied the coupling between the nuclei and the
electrons by measuring the nuclear spin relaxation time $T_{1}$, at various
positions near the crossing point C. First, rf was tuned into resonance, and
$R_{xx}$ shows a sharp decrease due to the nuclear depolarization. Then, the
frequency was switched back to off resonance. Nuclear spins that have once
flopped hardly relax back because of their longer relaxation time $T_{1}$,
which is on the order of minutes, relative to that of the electrons. Hence, $%
R_{xx}$ slowly relaxes back to its original value, and $T_{1}$ can be
derived by fitting $R_{xx}$ to the relation $R_{xx}=\alpha +\beta \exp
(-t/T_{1})$. Fig. \ref{C_NMR_T1} shows the data around point C to determine $%
T_{1}$.

Further insight is gained by investigating the NMR signals along the line L1
(please see Fig. \ref{NMR_Phase}a). As depicted in Fig. \ref{L1_NMR}a, our
measurement shows a clear peak of NMR ratio $\Delta R_{xx}/R_{xx}$ at the
crossing point C where the easy-axis pseudo-spin ferromagnetic states is
well developed. The obtained values of nuclear spin relaxation rate $1/T_{1}$
along line L1 are also plotted in Fig. \ref{L1_NMR}a. $1/T_{1}$ rapidly
increases from nearly zero to $8\times 10^{-3}$ (1/s) toward to the crossing
point C, as electron becomes the pseudo-spin ferromagnetic states. For
comparison, in Fig. \ref{L1_NMR}b we also show the electron activation
energy gap $E_{a}$ along the line L1. The single particle energy difference $%
E_{z}$ acts as effective Zeeman energy, and $E_{a}$ shows a slope of $5$
times greater than the single particle Zeeman gap $E_{z}$. This unusual
behavior is likely to be caused by the easy-axis ferromagnetism \cite%
{Jiang2006,Hirayama2001}. These quantities all show an obvious change as
approaching to the crossing point and demonstrate that $1/T_{1}$ is a
sensitive indicator of the pseudo-spin ferromagnetic formation. The
similarity between these phenomenon strongly suggest that an intimate link
between the spin and pseudo-spin in the easy-axis pseudo-spin ferromagnetic
states.

Interestingly, the data shown in Fig. \ref{L1_NMR}b shows that the slop of
activation energy gap $E_{a}$ to single particle Zeeman gap is as large as $%
5 $, which implies many spin flips within the magnetic domain walls and
support low energy mode of spin excitations \cite%
{Eisenstein1995,MacDonald2001}. As approaching to the crossing point C,
there are low energy spin excitations which give new channel to relax the
nuclear spin through the electron and nuclear spin flip-flop process. Thus
the NMR signal ratio $\Delta R_{xx}/R_{xx}$ and the nuclear spin relaxation
rate $1/T_{1}$ enhanced.

Despite the fact that the bulk of the results can be understood within the
framework of pseudo-spin quantum Hall ferromagnetism, there is still an
apparent puzzle. While we can find very strong NMR signals at the upper arm
of the square structure around point C, there is no detectible signal at the
lower arm of this square structure around point A. Since the two points have
equivalent LLs crossing configurations, one would expect that they are the
same easy-axis QHPF states and should produce similar NMR responses. In
principle, the NMR signal can be suppressed by spin-orbital coupling \cite%
{HirayamaNMR2} or mobility of domains \cite{HirayamaNMR3}. However, in our
case, point A and C have identical strength in spin-orbital coupling and
disorder. Therefore, the anomalous suppression of NMR signal at point A may
suggest that there could be some additional physics which has not yet been
recognized in the theory of pseudo-spin quantum Hall ferromagnetism.

In summary, RD-NMR has been measured in a two-subband electron system around
the LLs crossing points at total filling factor $\nu =3,5$ and $4$ where
easy-plane or easy-axis QHPF states are well developed. It reveals that the
easy-axis quantum Hall pseudospin state of $\nu =4$ is sensitive to the
RD-NMR measurement. As approaching to one LL crossing point at $\nu =4$, the
RD-NMR signal strength and the nuclear spin relaxation rate $1/T_{1}$
enhance quickly which may be due to the low energy spin excitations. At
another identical LL crossing point of $\nu =4$, the RD-NMR signal is found
to be suppressed and remains as a puzzle to be understood. Of course further
study is necessary to access the detailed mechanism.

This work at USTC was funded by National Basic Research Programme of China
(Grants No. 2006CB921900 and No. 2009CB929600), the Innovation funds from
Chinese Academy of Sciences, and National Natural Science Foundation of
China (Grants No. 10604052 and No. 10874163 and No.10804104). The work at
UCLA was supported by the NSF under Grant No. DMR-0804794.


\begin{thebibliography}{99}
\bibitem{DasSarma} Chap 2 and 5 in \textit{Perspectives on Quantum Hall
Effects, }S. Das Sarma and A. Pinczuk eds., (Wiley, New York, 1997).

\bibitem{Wescheider1999} V. Piazza, V. Pellegrini, F. Beltram, W.
Wegscheider, T. Jungwirth, and A. H. MacDonald, Nature \textbf{402}, 638
(1999).

\bibitem{Shayegan2000} E. P. De Portere, E. Tutuc, S. J. Papadakis, M.
Shayegan, Science \textbf{290}, 1546 (2000).

\bibitem{Hirayama2001} K. Muraki, T. Saku, and Y. Hirayama, Phys. Rev. Lett.
\textbf{87}, 196801 (2001).

\bibitem{Jiang2005} X. C. Zhang, D. R. Faulhaber and H. W. Jiang, Phys. Rev.
Lett. \textbf{95}, 216801 (2005).

\bibitem{Jiang2006} X. C. Zhang, I. Martin and H. W. Jiang, Phys. Rev. B
\textbf{74}, 073301 (2006).

\bibitem{Tsui2006} K. Lai, W. Pan, D.C. Tsui, S. Lyon, M. Muhlberger and F.
Schaffler, Phys. Rev. Lett. \textbf{96}, 076805 (2006).

\bibitem{Shayegan2006} K. Vakili, T. Gokmen, O. Gunawan, Y. P. Shkolnikov,
E. P. De Poortere and M. Shayegan, Phys. Rev. Lett. \textbf{97}. 116803
(2006).

\bibitem{MacDonald2000} T. Jungwirth and A. H. MacDonald, Phys. Rev. B
\textbf{63}, 035305 (2000).

\bibitem{DasSarma2003} D. W. Wang, E. Demler, and S. Das Sarma, Phys. Rev. B
\textbf{68}, 165303 (2003).

\bibitem{Hao2008} X. J. Hao et al., arXiv:0807.0297.

\bibitem{KlitzingNMR1} J. H. Smet et al., Nature (London) \textbf{415}, 281
(2002).

\bibitem{KlitzingNMR2} O. Stern et al., Phys. Rev. B \textbf{70}, 075318
(2004).

\bibitem{PortalNMR} W. Desrat et al., Phys. Rev. Lett. \textbf{88}, 256807
(2002).

\bibitem{TsuiNMR} G. Gervais et al., Phys. Rev. Lett. \textbf{94}, 196803
(2005).

\bibitem{EisensteinNMR} I. B. Spielman et al., Phys. Rev. Lett. \textbf{94},
076803 (2005).

\bibitem{HirayamaNMR1} N. Kumada et al., Phys. Rev. Lett. \textbf{94},
096802 (2005).

\bibitem{JiangNMR} X. C. Zhang, G. D. Scott and H. W. Jiang, Phys. Rev.
Lett. \textbf{98}, 246802 (2007).

\bibitem{MacDonald2001} T. Jungwirth and A. H. MacDonald, Phys. Rev. Lett.
\textbf{87}, 216801 (2001).

\bibitem{Eisenstein1995} A. Schmeller, J. P. Eisenstein, L. N. Pfeiffer, and
K. W. West, Phys. Rev. Lett. \textbf{75}, 4290 (1995).

\bibitem{HirayamaNMR2} K. Hashimoto et al., Phys. Rev. Lett. \textbf{94},
146601 (2005).

\bibitem{HirayamaNMR3} Y. Hirayama et al., Physica E. \textbf{20}, 133
(2003).
\end{thebibliography}
\end{document}